\def\lesssim{\,\lower2truept\hbox{${<\atop\hbox{\raise4truept\hbox{$\sim$}}}$}\,
}
\def\gtrsim{\,\lower2truept\hbox{${>\atop\hbox{\raise4truept\hbox{$\sim$}}}$}\,}

\def\deg {$^\circ$}                     
\documentclass{aa}
\input psfig.tex
\begin{document}
\thesaurus{11.05.1; 11.05.2; 11.06.1; 11.06.2; 11.09.2; 11.19.3}
\title{Star formation history of early--type galaxies in low density
environments.}

\subtitle{IV. What do we learn from  nuclear line-strength indices?}

\author{M. Longhetti \inst{1}, A. Bressan \inst{2}, C. Chiosi
\inst{3}\thanks{Visiting Scientist, Max-Planck Institut f\"ur Astrophysik,
K-Schwarzschild str. 1, D-87540, Garching bei M\"unchen, Germany},
R.  Rampazzo \inst{4}}

\offprints{M. Longhetti}

\institute{
$^1$Institut d'Astrophysique, 98 bis Boulevard Arago, 75014 Paris, France\\
$^2$Osservatorio Astronomico di Padova, Vicolo dell'Osservatorio,
5, 35122 Padova, Italy\\
$^3$Dipartimento di Astronomia, Universit\`a di Padova, Vicolo dell'Osservatorio,
5, 35122 Padova, Italy \\
$^4$Osservatorio Astronomico di Brera, Via Brera 28, 20121 Milano,
Italy }

\date{Received 8 March 1999; Accepted 31 August 199}

\authorrunning{Longhetti et al.}
\titlerunning{Star formation in early-type galaxies} 

\maketitle


\begin{abstract}
In this paper we analyze the line-strength indices in the 
Lick-system measured  by 
Longhetti et al. (1998a, b) for a sample of 51 
early-type galaxies located in low density environments
(LDE) and showing 
signatures of fine structures and/or interactions.
The sample contains 21 shell-galaxies and 30 members of interacting
pairs.

Firstly we perform a preliminary comparison between three
different sources of calibrations of the  line strength indices,
namely Buzzoni et al. (1992, 1994),  Worthey
(1992),  Worthey et al. (1994) and Idiart et al. (1995), 
derived from stars with 
different effective temperature, gravity,  and metallicity. 
Looking at the three indices in common, i.e. Mg2, Fe5270,
and H$\beta$, the calibrations by 
Buzzoni et al. (1992, 1994), Worthey (1992) and Worthey et
al. (1994) lead to mutually consistent results. 
The calibration of H$\beta$ by Idiart et al. (1995) can be compared
with the previous ones only for a limited range of ages,
in which  good agreement is found. 
Mg2 and Mgb indices predicted by the Idiart's et al. (1995) fitting functions
result to be systematically lower than those obtained from  using Worthey 
(1992) calibrations. 

Secondly, we discuss
the properties of the galaxies in our sample by comparing them both
with theoretical Single Stellar Populations (SSPs) and the {\sl normal} 
galaxies  of the Gonz\'alez (1993: G93) sample.
The analysis is performed by means of several diagnostic planes.

In the $\sigma$, Mg2, Fe5270 and Fe5335 space, {\sl normal}, 
shell- and pair-galaxies have a different
behavior.  First of all, normal and pair-galaxies follow the universal $\sigma$ vs.
Mg2 relation, whereas shell-galaxies lie above it; secondly the Fe versus Mg2 relation
of normal, shell- and pair-galaxies is flatter than the theoretical expectation. 
This fact hints for enhancement of $\alpha$-elements with respect to solar
partition in galaxies with strong Fe indices and/or high velocity dispersion,  
mass and  luminosity in turn.

In the $\sigma$ vs. H$\beta$ plane {\sl normal} galaxies seem to follow a nice relation
suggesting that objects with shallow gravitational potential  have strong H$\beta$
values (youth signature?), whereas shell- and pair-galaxies scatter all over the plane.
A group of galaxies with deep gravitational potential and strong H$\beta$
is found. Is this a signature of recent star formation?

In the H$\beta$ vs. [MgFe] plane,
\footnote{ $[MgFe] = \sqrt{ <Fe>\times Mgb}$, 

$\ <Fe>\ = (Fe5270 + Fe5335)/2$ }
which is perhaps best suited  
to infer the age of the stellar populations, 
the peculiar galaxies in our sample show nearly the same
distribution of the {\sl normal} galaxies in the 
G93 sample. 
There is however a number of peculiar galaxies with much stronger
H$\beta$. Does this mean that the scatter in the  
H$\beta$ vs. [MgFe] plane, of normal, shell- and pair-galaxies has a common
origin, perhaps a secondary episode of star formation? 
We suggest that, owing to their apparent {\sl youth}, shell- and pair-galaxies
should have experienced at least one
interaction event after their formation.
The explanation comes natural for shell- and pair-galaxies where the signatures
of interactions are evident. It is more intrigued in {\sl normal} galaxies
(perhaps other causes may concur).

Noteworthy, 
the distribution in the H$\beta$ vs. [MgFe] plane of {\sl normal}, shell- and 
pair-galaxies
is confined within a
narrow strip that runs  significantly steeper than the path followed by aging
SSPs.  
This feature is
explained as due to metal enrichment always accompanying star 
formation.

Shell-galaxies encompass the whole range of ages inferred from the 
H$\beta$ vs. [MgFe] plane, indicating that among them  recent and old
interaction/acquisition events are equally probable. 
If shells are formed at the same time at which the rejuvenating
event 
took place, shells ought to be long lasting phenomena.

\end{abstract}

\section{Introduction}

Hierarchical clustering scenarios predict that early-type galaxies we see today
in large virialized structures and in low density environments have formed at
different epochs (Ellis 1998 and references therein). Cluster objects
formed through major merging events at red-shift $z > 3$, whereas early-type galaxies,
today at the border of large structures, have formed significantly
later, at $z < 1$.

The morphology of early-type galaxies shows many relationships of
different kind with the surrounding   environment.
 The population of the early-type galaxies in clusters appears
to be homogeneous contrary-wise  to the large  heterogeneity shown  by
objects inhabiting the low density media. Although the bulk of 
galaxies visible today in low
density environments are already in place at $z \approx $ 0.5 (Griffith at al.
1994), i.e nearly half the Hubble time, a large number of them show 
peculiarities such as  signatures of interaction, fine structures etc. 
(see Schweizer 1992; Reduzzi et al. 1996
and references therein) which may hint at more recent activity.

In the above framework, early-type galaxies are currently interpreted as the
final product of  major/minor merging events (Schweizer 1992;
Barnes 1996). Nevertheless, both from observational and theoretical points
of view, evidence has grown over the years  that different
mechanisms have  also played a
significant role in shaping their final structure. In clusters, galaxy
encounters are fast enough to make mergers less probable than
{\sl harassments}   (Moore et al.  1996). In low density environments
encounters are most likely to result in a merger  (Barnes \& Hernquist
1992) but  dissipative mechanisms (Bender 1997) and {\sl weak-interaction}
events (Thomson  1991) could also contribute to the structural
evolution of  a galaxy.

This paper is the fourth of a series dedicated to the study of typical
galaxies in low density environments, i.e. galaxies showing signatures
of present/past interactions. The sample is composed of 21 shell-galaxies
and 30 members of interacting pairs most of which show fine
structures. Among them shell-galaxies represent a class of objects
exhibiting signatures of past interactions, i.e. minor/major mergers
(Schweizer  1992) or weak-interactions (Thomson \& Wright 1990) and
Thomson (1991). Pair-members are
instead objects with ongoing interaction.

Longhetti et al.  (1998a) measured 19
line-strength indices  in  the nuclear regions of the galaxies in our
sample.  Sixteen indices belong to the group  defined by 
Worthey (1992) and  G93 and include H$\beta$, Mg2
and some Fe features. All indices
were transformed into the Lick--IDS system. Furthermore, 
three  indices, particularly sensitive  to recent star formation
(Rose 1984, 1985; Leonardi \& Rose 1996),  i. e. $\Delta$4000,
H$\delta$/FeI and CaII(H+K), were added to the list.
Longhetti et al. (1998b) derived the inner kinematics of the sample and
corrected the line--strength
indices for central velocity dispersion (Longhetti et al. 1998a).

The comparison of these galaxies with those in Virgo and Fornax (Rampazzo et al.
1999a) singled out a group  of pair-galaxies
in our sample with  peculiar behavior in the  (log R$_e$, $\mu_e$)
plane and the $\kappa$ space. These galaxies, apparently missing among
cluster early-type  objects,  are  tidally stretched and most likely 
in  early stages of  interaction.  This finding makes evident
 the large scatter of line--strength indices as compared to galaxy
scale properties. As an example, the correlation between galaxy
shapes (as measured by the a4/a parameter) and  ages (as deduced from
H$\beta$)
that was advanced by   de Jong \& Davies
(1997) is not confirmed by the Rampazzo et al. (1999a,b) study.

The purpose of this  paper is to cast light on the past star formation
(SF) history of  shell- and pair-galaxies  in LDE  
as traced  by their line strength indices,
with particular attention to the role played by
dynamical interactions in triggering star formation. The  comparison with
the results obtained for galaxies in denser environments (Burstein
et al. 1984; Pickles 1985; Rose 1995; Bower et al. 1990;
G93)  will  help us to understand the effects of the environment
on   the formation/evolution of early-type galaxies.

In addition to this, we address the problem of the origin of
shell structures by analyzing the evolutionary history of the stellar populations
hosted by the nucleus of the interacting  galaxy.  Numerical simulations of dynamical
interactions among galaxies yield still  contrasting explanations
for the occurrence of shells (Weil \& Hernquist 1993; Thomson 1991).
In this study, we seek to assess  the duration of the shell
phenomenon by means of    the age of the last episode of star
formation as  inferred from the nuclear indices.

The paper is organized as follows. In Sect.~2 we compare indices for
Single Stellar Populations (SSPs) obtained  using
{\sl fitting functions} by different authors (Buzzoni et al. 1992, 1994;  
Worthey 1992; Idiart et al. 1995) and  a unique source of
isochrones (Bertelli et al. 1994). By doing so, we are able to
quantify the 
uncertainties affecting line strength indices calculations. 
Based on this preliminary comparison, we  adopt the fitting functions of  
Worthey (1992). In Sect.~3 we introduce the sample
of galaxies of G93 adopted as the reference template. 
In Sect.~4, firstly we compare the observational
 line strength indices for the nuclear region of galaxies in our
sample with the theoretical predictions, and secondly we 
compare our sample of shell- and pair-galaxies with that of  
{\sl normal} elliptical galaxies by G93.
Notes on individual galaxies in the H$\beta$ vs. [MgFe] diagram
are reported in Sect.~5. A tentative explanation of the distribution
of  galaxies in the H$\beta$ vs. [MgFe] diagram is presented in
Sect.~6  both for the present sample and the template. The
interpretation is based on  statistical  simulations of the
observations. Finally,  
 Sect.~7 summarizes the results.

\begin{table*}
\caption[1]{Comparison between SSP indices derived from different fitting functions}
\begin{center}
\begin{tabular}{rrr rrr rrr rr}
\noalign{\smallskip}
\hline
\noalign{\smallskip}
  &  $\Delta$\%  &  $\sqrt{\Delta^2}$\%  &  $\Delta$\%  &
$\sqrt{\Delta^2}$\%  &  $\Delta$\%
  &  $\sqrt{\Delta^2}$\%  &  $\Delta$\%  &  $\sqrt{\Delta^2}$\%  &
$\Delta$\%  & $\sqrt{\Delta^2}$\%  \\
  &    &    &  T$>$3Gyr  &  T$>$3Gyr  &  T$<$3Gyr  &  T$<$3Gyr  &
T$>$1Gyr  &  T$>$1Gyr  &  T$<$1Gyr  &  T$<$1Gyr  \\
\noalign{\smallskip}
\hline
\noalign{\smallskip}    
\multispan{2}{Worthey - Idiart}   &    &    &    &    &    &    &
&    &    \\
\noalign{\smallskip}
\hline
H$\beta$&         &        & -34.72 & 40.36 &         &        &
       &        &           &       \\
Mg2     &   12.72 &  12.94 &  9.45  &  9.45 &  22.97  &  23.53 &
10.31  &  10.32  &  27.28  & 6.92  \\
Mgb     &   15.50 &  15.76 &  11.19 & 11.23 &  29.17  &  29.87 &
12.39  &  12.44  &  34.65  & 8.76  \\
\noalign{\smallskip}
\hline
\noalign{\smallskip}
\multispan{2}{Worthey - Buzzoni}  &    &    &    &    &    &    &
&    &    \\
\noalign{\smallskip}
\hline
H$\beta$&  -6.34  &  11.19 &  -4.29 &  4.75 &  -6.99  & 10.31  &
-2.51  &  3.85  &  -8.40  & 2.64  \\
Mg2     &  96.88  & 168.29 &  -1.82 &  3.20 &-837.16  &-1006.74&
2.75  & 11.27  &  -386.13  & -107.19  \\
Fe5270  &  36.49  &  76.69 &  -7.89 &  8.21 &  284.63 &  354.45&
-5.11  &  9.09  &  1837.04  & 520.67  \\
Fe5335  &   6.52  &  10.54 &  -0.80 &  2.65 &  24.30  &  24.74 &
0.39  &  5.79  &  28.63  & 29.23  \\
\noalign{\smallskip}
\hline
\end{tabular}
\end{center}
\label{t1}
\end{table*}

\begin{table}
\caption[2]{SSP indices obtained adopting the Lick fitting functions }
\begin{scriptsize}
\begin{center}
\begin{tabular*}{70mm}{l rrr rrr }
\noalign{\medskip}
\hline
\noalign{\medskip}
\multicolumn{7}{c}{Z=0.004}\\               
\noalign{\medskip}
\hline
\noalign{\medskip}
   Age & H$\beta$ & Mg2  &  Mgb & Fe52 & Fe53 & MgFe \\  
\noalign{\medskip}
\hline
\noalign{\medskip}
  7.70   & 5.10 & -0.57 & 0.44 & 0.64 & 0.44 & 0.49  \\
  7.85   & 5.51 & -0.58 & 0.45 & 0.52 & 0.33 & 0.44  \\      
  8.00   & 5.72 & -0.56 & 0.49 & 0.48 & 0.28 & 0.43  \\  
  8.95   & 5.09 &  0.06 & 0.97 & 0.83 & 0.61 & 0.84  \\
  9.00   & 4.82 &  0.06 & 1.05 & 0.93 & 0.71 & 0.93  \\
  9.18   & 4.01 &  0.08 & 1.27 & 1.27 & 0.96 & 1.19  \\
  9.70   & 2.41 &  0.12 & 1.94 & 1.77 & 1.47 & 1.77  \\ 
  9.78   & 2.27 &  0.13 & 2.03 & 1.83 & 1.53 & 1.85  \\ 
  9.85   & 2.14 &  0.13 & 2.12 & 1.89 & 1.59 & 1.92  \\ 
  9.90   & 2.02 &  0.14 & 2.19 & 1.95 & 1.64 & 1.98  \\ 
  9.95   & 1.98 &  0.14 & 2.24 & 1.95 & 1.65 & 2.01  \\ 
  10.00  & 1.91 &  0.14 & 2.28 & 1.99 & 1.69 & 2.05  \\ 
  10.04  & 1.84 &  0.15 & 2.33 & 2.02 & 1.74 & 2.09  \\  
  10.08  & 1.78 &  0.15 & 2.37 & 2.06 & 1.78 & 2.14  \\  
  10.18  & 1.71 &  0.15 & 2.42 & 2.06 & 1.81 & 2.16  \\
\noalign{\medskip}
\hline  
\noalign{\medskip}
\multicolumn{7}{c}{Z=0.02}\\
\noalign{\medskip}
\hline
\noalign{\medskip}
   7.70  & 5.25 & 0.06 & 0.61 & 0.88 & 0.91 & 0.74 \\
   7.85  & 5.65 & 0.05 & 0.65 & 0.84 & 0.87 & 0.75 \\
   8.00  & 6.00 & 0.05 & 0.73 & 0.81 & 0.86 & 0.78 \\
   8.95  & 4.44 & 0.10 & 1.65 & 1.78 & 1.54 & 1.65 \\
   9.00  & 4.06 & 0.12 & 1.82 & 1.94 & 1.69 & 1.82 \\
   9.18  & 3.31 & 0.15 & 2.25 & 2.24 & 1.99 & 2.18 \\
   9.70  & 1.94 & 0.22 & 3.37 & 2.84 & 2.54 & 3.01 \\
   9.78  & 1.83 & 0.23 & 3.49 & 2.90 & 2.60 & 3.10 \\
   9.85  & 1.76 & 0.23 & 3.56 & 2.92 & 2.61 & 3.14 \\
   9.90  & 1.68 & 0.24 & 3.66 & 2.97 & 2.66 & 3.21 \\
   9.95  & 1.61 & 0.25 & 3.74 & 3.02 & 2.71 & 3.27 \\
 10.00   & 1.54 & 0.25 & 3.83 & 3.07 & 2.75 & 3.34 \\
 10.04   & 1.49 & 0.26 & 3.91 & 3.12 & 2.80 & 3.40 \\ 
 10.08   & 1.45 & 0.26 & 3.96 & 3.15 & 2.83 & 3.44 \\
 10.18   & 1.37 & 0.27 & 4.08 & 3.21 & 2.89 & 3.53 \\
\noalign{\medskip}
\hline
\noalign{\medskip}
\multicolumn{7}{c}{Z=0.05} \\
\noalign{\medskip}
\hline
\noalign{\medskip}
   7.70  & 6.21 & 0.06 & 0.75 & 0.85 & 1.14 & 0.86 \\
   7.85  & 6.46 & 0.06 & 0.75 & 0.92 & 1.21 & 0.90 \\
   8.00  & 6.74 & 0.06 & 0.75 & 0.96 & 1.25 & 0.91 \\
   8.95  & 3.81 & 0.16 & 2.30 & 2.57 & 2.36 & 2.38 \\
   9.00  & 3.50 & 0.17 & 2.50 & 2.71 & 2.48 & 2.55 \\
   9.18  & 2.79 & 0.21 & 3.02 & 3.03 & 2.78 & 2.96 \\
   9.70  & 1.70 & 0.30 & 4.34 & 3.60 & 3.31 & 3.87 \\
   9.78  & 1.60 & 0.31 & 4.51 & 3.67 & 3.38 & 3.99 \\
   9.85  & 1.50 & 0.33 & 4.67 & 3.75 & 3.47 & 4.11 \\
   9.90  & 1.47 & 0.33 & 4.72 & 3.76 & 3.47 & 4.13 \\
   9.95  & 1.41 & 0.34 & 4.84 & 3.81 & 3.52 & 4.21 \\
  10.00  & 1.36 & 0.34 & 4.94 & 3.86 & 3.57 & 4.28 \\
  10.04  & 1.34 & 0.35 & 4.95 & 3.87 & 3.58 & 4.29 \\
  10.08  & 1.32 & 0.35 & 4.99 & 3.88 & 3.58 & 4.32 \\
  10.18  & 1.24 & 0.36 & 5.09 & 3.96 & 3.66 & 4.41 \\
\noalign{\medskip}
\hline
\end{tabular*}
\end{center}
\end{scriptsize}
\label{t2}
\end{table}

\begin{figure}
\psfig{file=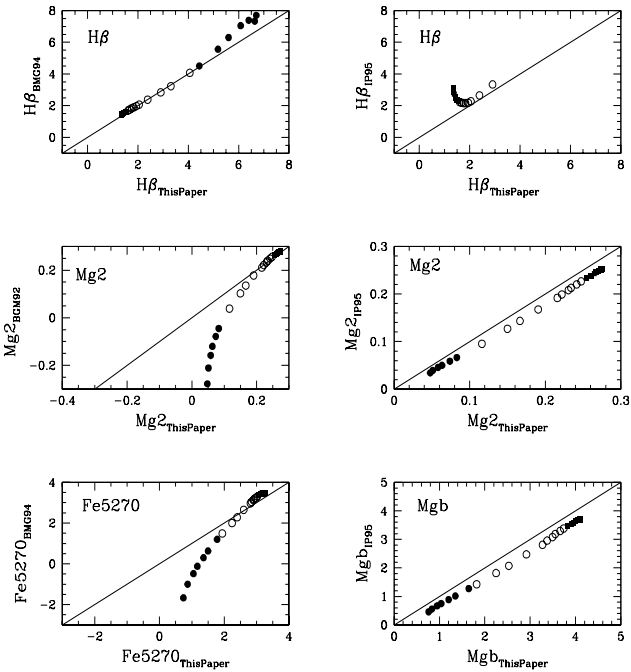,height=9.0truecm,width=8.5truecm} 
\caption{Comparison between indices  derived from the Lick-system fitting
functions (adopted in this paper)
and those from  Buzzoni et al. (1992, 1994; 
shortly indicated by  BGM92 and BMG94 respectively) and
Idiart et al. (1995: IP95).  In the case of 
Buzzoni et al. (1992, 1994) we  compare  the indices
H$\beta$, Mg2
and Fe5270. In the case of Idiart et al. (1995) we examine the indices
H$\beta$,
Mg2 and Mgb. The data refer to SSPs with
solar metallicity (Z = 0.02) and different ages:
full circles for ages between 0.2 and 0.9 Gyr;
open circles for ages between 1 and 9; full squares for ages between 10 and 19
Gyr.}
\label{fig_bu_wo_id}
\end{figure}

\section{Line-strength indices for SSPs}

The technique to calculate  line-strength indices of SSPs is amply
described in Worthey (1994),  Bressan et al. (1996) and 
Tantalo et al. (1998). The latter improved their previous models
by introducing a differential method particularly suited to
understand the role of the three main parameters driving the strength
of selected indices, namely age, metallicity and enhancement 
in the abundance of  $\alpha$
elements  with respect to the solar partition.
The reader is referred to those articles and references therein for a
detailed description of the method. Here we limit ourselves to
summarize a few basic
properties of the models  in usage.
The galaxy indices  are based on the isochrones of Bertelli et
al. (1994),
which provide the relative number of stars per elemental area of the
Hertzsprung-Russell Diagram (HRD), 
the atlas of theoretical stellar spectra by Kurucz (1992), which is used  
to calculate the energy distribution in the continuum of each
pass-band of interest, and a set of fitting functions. 

The elemental areas of the HRD along the path drawn by an isochrone
are taken sufficiently small so that all stars in each of them have the
same effective temperature, gravity, luminosity and chemical
composition.

The Kurucz (1992) library is extended at high temperatures by black-bodies 
and at temperatures
cooler than 3500\deg\ K  (late K and M type) by  implementing 
the  libraries of Lancon \& Rocca-Volmerange (1992),  
Straizys \& Sviderskiene (1972), and Fluks et al. (1994), see Bressan
et al. (1994) and Tantalo et al. (1996) for all the details.

The {\sl fitting functions} depend on stellar effective temperature, 
gravity and metallicity.
They are used to calculate the line strength indices for each
elemental
area of the HRD. Finally a  suitable integration technique
yields the total indices for each SSP (isochrone).

Direct measurements of line strength indices on the integrated spectral
energy distribution of individual SSPs is not possible, 
because the Kurucz (1992) library contains spectra at low 
resolution.
A different strategy would be to replace this library 
with another one containing medium resolution spectra on
which the direct measurement of the indices could be made.
This possibility, explored in a forthcoming paper, is no
longer considered here.

\subsection{Uncertainties on the  line-strength indices}

In this section  we compare line-strength indices computed by adopting
fitting functions from different authors, namely Worthey (1992) and
Worthey et al.
(1994), otherwise known as the Lick-system,  Buzzoni et al. (1992, 1994), and 
Idiart et al. (1995).

The Lick fitting functions refer to 21 line--strength indices (Worthey 1992;
Gorgas et al. 1993; Worthey et al. 1994) and are based on a library of 
stellar spectra containing  about 400 stars, observed at the Lick
Observatory  between 1972 and
1984, with an Image Dissector Scanner (IDS). 
In the following  we adopt the Worthey (1992) fitting functions, 
extended however to high temperature stars ($T_{eff} \approx$
10000\deg K)  as reported in Longhetti et al. (1998a).

The Buzzoni et al. (1992) fitting functions for the Mg2, Fe5270 and H$\beta$
indices, rest on a library of spectra for 74 stars. Buzzoni et
al. (1994) do not
consider the dependence of the H$\beta$  on  the metallicity.

Idiart et al. (1995) present  calibrations for  Mg2, Mgb and H$\beta$,
based  on a library  of 170 stars, among which 89 are new observations, and the
remaining are from Faber et al. (1985) and Gorgas et al. (1993) data. The 
sample spans the metallicity range  $-3.0 < $[Fe/H]$ < 0.2$, the  gravity range 
$0.7 < log(g) <5.0$  and the  temperature range  $3800$\deg K $< T_{eff} <
6500$\deg K.

We construct integrated narrow band indices for SSPs of solar metallicity
by applying the different fitting functions above  to the same set of
stellar  isochrones.
This allows us to single out the effects of different empirical relations
and evaluate  the corresponding uncertainty.

The comparison between the three different sets of models for the indices
in common is shown in
Fig. \ref{fig_bu_wo_id}.

Good agreement between Worthey (1992) and  Buzzoni et al. (1992,
1994) calibrations 
exists for the H$\beta$ index of SSPs older than 1 Gyr, where the mean 
difference
between the two sources is $\approx 2$\% (see Table \ref{t1}).  For SSPs younger
than 1 Gyr the  fitting functions by Buzzoni et al. (1992, 1994) slightly 
overestimate the index H$\beta$ with respect to those by Worthey (1992).
Furthermore, the Mg2 and Fe5270
indices,  more sensitive to changes
in  metallicity, agree  only for SSPs older than 3 Gyr. For younger SSPs, 
they differ by large factors. 
It is worth recalling that Buzzoni et al. (1992, 1994) considered
their calibration to be applicable only to old SSPs.

\noindent
The calibrations  by  Idiart et al. (1995) fit the behavior of
Mg2, Mgb and H$\beta$ for stellar temperatures between
3800 K and 6500 K.  The features measured by the
Mg2 and Mgb indices are strongly dominated by relatively cool stars,
both in young and  old SSPs. On the contrary, the H$\beta$
index  mainly reflects 
the contribution of main sequence and turn off stars of
a stellar population. 

Unfortunately, at ages younger than 2 Gyr the effective 
temperatures of these stars are higher than 6500 K, the upper
limit of those fitting functions (Bertelli et al 1994). 
Therefore, values obtained for the
H$\beta$ index adopting the calibration of Idiart et al. (1995)
are reliable only for relatively old SSPs.
Indeed, when   only  SSPs older than
2 Gyr are compared, good agreement
between  Idiart's et al. (1995) and Worthey's et al. (1994)
  calibration of H$\beta$ is found (see Fig. \ref{fig_bu_wo_id}). 
In contrast, for  ages older than 10 Gyr,
the effect of main sequence stars cooler than 3800 K (for which
the index has been extrapolated out of the range of
validity of the calibrations) can be seen in the growing
disagreement between the predictions of Idiart et al. (1995)
and those of the Lick system.

The predictions for Mg2 and Mgb  do not suffer from this
limitation in temperature, therefore they can be extended also to 
young SSPs as shown in Fig. \ref{fig_bu_wo_id}. The 
Mg2 and Mgb indices derived from  Idiart's et al. (1995) fitting functions
are systematically lower than those  from Worthey (1992)
calibrations, with an offset of 0.02 mag for Mg2 and  0.36 mag for Mgb.
These systematic shifts are probably due to the different  sample of stars
adopted by Idiart et al. (1995), that leads to a different 
calibration of the indices with respect to the stellar gravity.

In Table \ref{t1} we present the detailed comparison of results obtained
from using Worthey (1992),
Buzzoni et al. (1992, 1994), and Idiart et al. (1995):  Table \ref{t1} lists 
$\Delta$\% (average difference between the two
sets of indices values) and $\sqrt{\Delta^2}$\% (root square of
the average quadratic difference) for different values of the age as indicated.

In the following we will adopt Worthey (1992) as the reference calibration,
and consider the average offset between
Buzzoni et al. (1992)  and Worthey (1992) predictions, as representative of the
uncertainty of the models. Indices calculated with this set of fitting functions
are reported in Table \ref{t2} for three different metallicities, and for
ages between $5\times 10^7$ yr and 15 Gyr.

\begin{figure}
\psfig{file=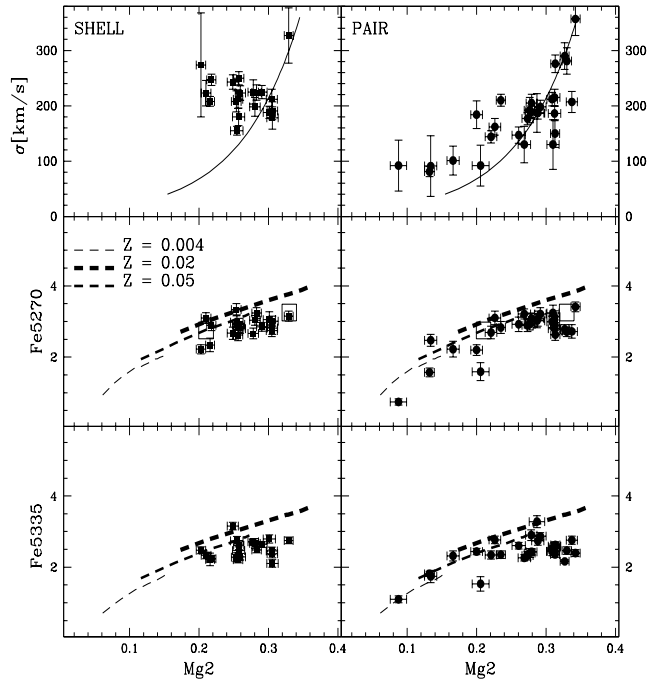,height=10.0truecm,width=8.8truecm}
\caption{Velocity dispersion $\sigma$ and iron indices (Fe5270 and Fe5335) vs.
Mg2 for shell- (left panels) and pair-galaxies (right panels). 
Lines in the upper panels represent the Mg2 - $\sigma$ relation
by Bender at al. (1993). The
dashed lines  represent the SSP models for three different
metallicities, namely  Z=0.004, Z=0.02 (the solar value), and Z=0.05
in the age range  1 to 15 Gyr (the age increases from  left  to  right).  
The open squares in the Fe5270 vs. Mg2 panels
represent the mean values calculated by Worthey et al. (1992)
for a sample of {\sl compact} (left)  and {\sl giant} galaxies (right) }
\label{fig_fe_sig}
\end{figure}

\section{ Remarks on the observational data}

\subsection{The reference sample by G93}
G93 obtained long-slit spectroscopic data for a sample of 41
elliptical galaxies. He derived kinematic profiles
and line-strength indices for the nuclear regions (within 1/8 of the
effective radius) and for a wider coverage of the galaxies area (within
1/2 of the effective radius), thus providing  information on 
the radial gradients in these quantities. G93 sample of galaxies is the reference frame
to which often in the course of this paper our data will be compared.
In order to make the G93 data fully consistent with our ones, we 
start from his raw data within the central 5\arcsec, to which we apply the 
correction for
velocity dispersion (see below). The corrected data are listed in 
Table \ref{t3}.

\begin{table*}
\caption[3]{The G93 data within 5\arcsec\ after correction for
velocity dispersion  $\sigma$ (in km/s).}
\begin{scriptsize}
\begin{center}
\begin{tabular*}{170mm}{lrr rrr rr  lrr rrr rr}
\hline
\noalign{\smallskip}
 Name   & H$\beta$ & Mg2  &  Mgb  &  Fe52 &  Fe53  & MgFe & $\sigma$& 
       Name   & H$\beta$ & Mg2  &  Mgb  &  Fe52 &  Fe53  & MgFe & $\sigma$  \\
\hline
\noalign{\smallskip}
NGC221  &  2.28 & 0.22  & 2.98 & 3.00  & 2.68 & 3.60 &  77 & NGC5812 & 1.71 & 0.34 & 
4.47 & 2.84 & 2.51 & 4.27&  200\\
NGC224  &  1.67 & 0.34  & 4.58 & 3.07  & 2.64 & 4.48 & 183 & NGC5813 & 1.23 & 0.32 & 
4.53 & 2.60 & 2.03 & 4.05&  220\\
NGC315  &  1.00 & 0.32  & 4.21 & 2.37  & 1.84 & 3.72 & 310 & NGC5831 & 1.83 & 0.30 & 
4.31 & 3.06 & 2.60 & 4.33&  163\\
NGC507  &  1.67 & 0.32  & 4.02 & 2.54  & 2.07 & 3.79 & 275 & NGC5846 & 1.17 & 0.34 & 
4.61 & 2.65 & 2.18 & 4.15&  228\\
NGC547  &  1.27 & 0.32  & 4.59 & 2.57  & 1.98 & 4.05 & 242 & NGC6127 & 1.52 & 0.33 & 
4.47 & 2.55 & 2.10 & 4.01&  245\\
NGC584  &  1.82 & 0.29  & 4.05 & 2.78  & 2.30 & 3.99 & 198 & NGC6702 & 2.10 & 0.25 & 
3.57 & 2.87 & 2.46 & 3.83&  174\\
NGC636  &  1.86 & 0.28  & 3.96 & 2.97  & 2.48 & 4.08 & 163 & NGC6703 & 1.63 & 0.29 & 
4.10 & 2.82 & 2.36 & 4.05&  184\\
NGC720  &  1.54 & 0.35  & 4.71 & 2.62  & 2.21 & 4.19 & 247 & NGC7052 & 0.89 & 0.33 & 
4.47 & 2.48 & 1.95 & 3.93&  286\\
NGC821  &  1.66 & 0.33  & 4.45 & 2.88  & 2.35 & 4.25 & 196 & NGC7454 & 2.05 & 0.23 & 
3.23 & 2.56 & 2.21 & 3.45&  103\\
NGC1453 &  0.89 & 0.31  & 4.21 & 2.52  & 1.94 & 3.83 & 294 & NGC7562 & 1.62 & 0.30 & 
4.07 & 2.67 & 2.03 & 3.87&  244\\
NGC1600 &  1.42 & 0.34  & 4.38 & 2.38  & 2.02 & 3.86 & 325 & NGC7619 & 1.46 & 0.34 & 
4.35 & 2.57 & 2.00 & 3.94&  314\\
NGC1700 &  1.92 & 0.29  & 3.77 & 2.81  & 2.23 & 3.85 & 226 & NGC7626 & 1.48 & 0.35 & 
4.59 & 2.13 & 2.06 & 3.80&  259\\
NGC2300 &  1.70 & 0.35  & 4.45 & 2.67  & 2.04 & 4.05 & 264 & NGC7785 & 1.48 & 0.31 & 
4.20 & 2.60 & 2.18 & 3.94&  239\\
NGC2778 &  1.29 & 0.35  & 4.43 & 2.81  & 2.35 & 4.20 & 157 & NGC4261 & 1.09 & 0.35 & 
4.52 & 2.68 & 1.96 & 4.07&  305\\
NGC3377 &  1.85 & 0.29  & 4.07 & 2.66  & 2.28 & 3.93 & 126 & NGC4374 & 1.00 & 0.33 & 
4.31 & 2.51 & 1.85 & 3.84&  289\\
NGC3379 &  1.42 & 0.34  & 4.47 & 2.73  & 2.21 & 4.14 & 212 & NGC4472 & 1.73 & 0.32 & 
4.33 & 2.51 & 2.14 & 3.94&  290\\
NGC3608 &  1.61 & 0.33  & 4.36 & 2.70  & 2.31 & 4.10 & 187 & NGC4478 & 1.81 & 0.29 & 
4.09 & 2.80 & 2.51 & 4.08&  127\\
NGC3818 &  1.48 & 0.33  & 4.60 & 2.78  & 2.42 & 4.28 & 174 & NGC4552 & 1.31 & 0.36 & 
4.69 & 2.63 & 2.21 & 4.19&  256\\
NGC4278 & -1.13 & 0.33  & 4.57 & 2.39  & 1.99 & 3.93 & 246 & NGC4649 & 1.38 & 0.37 & 
4.55 & 2.44 & 1.88 & 3.92&  349\\
NGC4489 &  2.30 & 0.25  & 3.27 & 2.98  & 2.49 & 3.72 &  48 & NGC4697 & 1.58 & 0.30 & 
4.32 & 3.00 & 2.54 & 4.29&  161\\
NGC5638 &  1.62 & 0.32  & 4.53 & 2.89  & 2.36 & 4.29 & 153 &         &      &      &    
  &      &      &     &     \\
\noalign{\smallskip}
\hline
\end{tabular*}
\end{center}
\end{scriptsize}
\label{t3}
\end{table*}

\subsection{Correcting G93 for velocity dispersion}

The correction of the G93 data for velocity dispersion
is made using the method of Longhetti et al. (1998a)
for the sake of internal consistency.
We remind the reader that G93 corrected his data for velocity
dispersion, but using a different method. Although Longhetti et al. (1998b)
demonstrated that good agreement exists between our  kinematic
parameters and those of G93,   
and that the differences brought about by the two methods for 
correcting indices for
velocity dispersion are 
statistically small, yet they may lead to systematic discrepancies 
between the two
data sets in  the  [MgFe] - H$\beta$ diagram. 
Since our aim is to single out physical differences between
{\sl normal} (G93) and  {\sl interacting or post-interacting}
galaxies, we need to clean the data for all possible
effects of spurious nature.

\subsection{ Effects of emission lines on G93 data }

Although the G93 sample was selected to exclude galaxies with large 
amounts of gas,   a substantial fraction of the ellipticals
in this list (28/40)  clearly shows evidence of emission (EW$\le$
0.15\AA) 
in the [OIII](5007\AA) line, at least in the central parts  of the galaxies.
Emission in not strong (only one galaxy, NGC 4278 has
[OIII](5007\AA) emission larger than 1\AA) but it can seriously affect the
interpretation of the H$\beta$ index. To cope with this difficulty,
G93 tried
to cure the H$\beta$ index of his  galaxies  for  emission contamination  
adopting a linear relation between the intensity of
the H$\beta$
emission line  and that of the forbidden line [OIII](5007\AA). This
correlation has been found empirically, measuring the two emission
lines on a sample of galaxies. However, the measure of the H$\beta$
emission line is not an easy task, because of the possible
contamination by the  {\sl absorption} component, and the correct
evaluation of the effect of H$\beta$ emission line on the
H$\beta$ index requires detailed models of the
emission/absorption  processes. Therefore, instead of applying 
uncertain corrections, we prefer to leave   H$\beta$ unchanged and
always keep in mind that G93 data adopted in the present work
are not corrected
for this effect.

\subsection{The sample of shell- and pair-galaxies }

All the data in our sample refer  to the central 5\arcsec\ portion of
the objects and  therefore are homogeneous  with those by
G93. Nevertheless,
we have checked how the indices would change when
5\arcsec\ region in our data is smaller or greater than 
1/8 of the effective radius. 
No sizable difference has been noticed, so that no
correction for aperture is applied.

Furthermore, all observational indices  
have been corrected for velocity dispersion
($\sigma$). This strictly follows the same procedure  as in 
Longhetti et
al. (1998a) so that no details are given here.
 
Finally, Table \ref{t4} lists 
all the data for our sample that will be used in the analysis below.

\begin{table*}
\caption[4]{Basic data for our sample of shell- and pair-galaxies. The
velocity dispersion $\sigma$ is km/s.}
\begin{scriptsize}
\begin{center} 
\begin{tabular*}{170mm}{lrr rrr rr  lrr rrr rr}
\hline
\noalign{\smallskip}
\multicolumn{8}{c}{Pair-Galaxies}&\multicolumn{8}{c}{Shell-Galaxies}\\
\hline
\noalign{\smallskip}
 Name   & H$\beta$ & Mg2  &  Mgb  &  Fe52 &  Fe53  & MgFe & $\sigma$& 
    Name&  H$\beta$ & Mg2  &  Mgb  &  Fe52 &  Fe53  & MgFe & $\sigma$ \\
\hline
\noalign{\smallskip}
 RR24a  & -2.60 & 0.17 & 2.49  & 2.22  & 2.32 & 2.90 &101  &
 N813   & 2.29 &  0.25 & 4.19 &  3.33 &  2.28 &  4.33 &208\\
 RR24b  &-17.37 & 0.09 & 2.01  & 0.74  & 1.10 & 1.61 & 92  &
 N1210  & 1.44 &  0.26 & 3.94 &  2.92 &  2.46 &  4.04 &180\\
 RR62a  &  1.15 & 0.13 & 2.30  & 1.57  & 1.82 & 2.39 & 81  &
 N1316  & 2.01 &  0.26 & 4.02 &  2.84 &  2.62 &  4.08 &250\\
 RR101a &  1.86 & 0.23 & 3.53  & 3.10  & 2.76 & 3.98 &162  &
 N1549  & 1.71 &  0.29 & 4.46 &  2.85 &  2.64 &  4.31 &225\\
 RR101b &  2.35 & 0.23 & 3.67  & 2.83  & 2.35 & 3.83 &210  &
 N1553  & 1.40 &  0.30 & 4.45 &  3.05 &  2.79 &  4.45 &188\\
 RR105a &  1.32 & 0.27 & 4.23  & 3.21  & 2.26 & 4.29 &130  &
 N1571  & 1.66 &  0.28 & 4.31 &  3.24 &  2.48 &  4.39 &224\\
 RR187b &  2.87 & 0.22 & 3.40  & 2.69  & 2.34 & 3.62 &144  &
 N2865  & 3.12 &  0.22 & 3.28 &  2.34 &  2.22 &  3.36 &208\\
 RR210a &  0.80 & 0.33 & 4.97  & 2.71  & 2.47 & 4.43 &281  &
 N2945  & 0.38 &  0.26 & 4.80 &  2.62 &  2.21 &  4.23 &212\\
 RR210b &  1.50 & 0.31 & 4.62  & 2.99  & 2.44 & 4.41 &212  &
 N3051  & 1.15 &  0.30 & 4.82 &  2.74 &  2.45 &  4.37 &212\\
 RR225a &  1.76 & 0.34 & 4.96  & 3.41  & 2.40 & 4.78 &357  &
 N5018  & 2.68 &  0.22 & 3.31 &  2.89 &  2.24 &  3.64 &247\\
 RR225b &  1.18 & 0.31 & 4.42  & 2.80  & 2.59 & 4.26 &186  &
 N6776  & 1.92 &  0.25 & 4.39 &  2.67 &  3.15 &  4.32 &243\\
 RR278a & -1.67 & 0.20 & 4.14  & 2.20  & 2.44 & 3.76 &184  &
 N6849  & 1.32 &  0.28 & 4.34 &  3.03 &  2.67 &  4.35 &198\\
 RR282b &  1.53 & 0.33 & 5.17  & 2.78  & 2.16 & 4.47 &290  &
 N6958  & 1.66 &  0.26 & 4.08 &  2.80 &  2.29 &  4.01 &223\\
 RR287a &  1.61 & 0.28 & 4.57  & 3.08  & 2.43 & 4.43 &205  &
 E1070040& 2.30 &  0.25 & 3.97 &  2.98 &  2.74 &  4.16 &155\\
 RR297a &  1.97 & 0.27 & 4.08  & 2.89  & 2.32 & 4.06 &177  &
 E3420390& 1.32 &  0.33 & 4.77 &  3.15 &  2.75 &  4.65 &327 \\
 RR297b &  1.17 & 0.26 & 4.81  & 2.92  & 2.60 & 4.51 &147  &
 N7135  &-0.41 &  0.31 & 5.51 &  2.81 &  2.38 &  4.70 &191\\
 RR298b &  1.42 & 0.31 & 4.48  & 3.24  & 2.44 & 4.47 &130  &
 E2890150& 0.29 &  0.20 & 3.06 &  2.22 &  2.47 &  3.25 &274\\
 RR307a &  0.08 & 0.21 & 3.73  & 1.59  & 1.53 & 2.97 & 92  &
 E2400100a  & 1.54 &  0.28 & 4.63 &  2.66 &  2.71 &  4.31 &225\\
 RR317a &  1.56 & 0.28 & 4.36  & 3.06  & 2.90 & 4.43 &189  &
 E2400100b  & 2.79 &  0.21 & 3.79 &  3.10 &  2.33 &  4.02 &223\\
 RR317b &  2.33 & 0.13 & 1.80  & 2.47  & 1.74 & 2.45 & 91  &
 E5380100   & 1.70 &  0.31 & 4.20 &  3.01 &  2.10 &  4.13 &178\\
 RR381a &  1.64 & 0.31 & 5.01  & 2.63  & 2.62 & 4.44 &276  & 
       &      &       &      &       &       &       &   \\
 RR387a &  2.20 & 0.29 & 4.48  & 3.08  & 2.74 & 4.47 &187  & 
       &      &       &      &       &       &       &   \\
 RR387b &  1.83 & 0.31 & 4.51  & 3.13  & 2.55 & 4.45 &215  & 
       &      &       &      &       &       &       &   \\
 RR397b &  1.33 & 0.29 & 4.39  & 3.21  & 2.87 & 4.52 &198  &
        &      &       &      &       &       &       &   \\
 RR405a &  1.45 & 0.28 & 4.68  & 2.92  & 2.39 & 4.39 &192  &
        &      &       &      &       &       &       &   \\
 RR405b &  1.89 & 0.34 & 4.81  & 2.72  & 2.76 & 4.44 &207  &
        &      &       &      &       &       &       &   \\
 RR409a &  1.89 & 0.29 & 4.51  & 3.02  & 3.28 & 4.59 &187  &
        &      &       &      &       &       &       &   \\
 RR409b &  1.17 & 0.31 & 4.56  & 2.89  & 2.36 & 4.31 &150  & 
       &      &       &      &       &       &       &   \\
\noalign{\smallskip}
\hline
\end{tabular*}
\end{center}
\end{scriptsize}
\label{t4}
\end{table*}

\section{Galaxy indices: theory versus data}

In this section we compare the observational indices of the galaxies in
our  sample with those from model calculations.

\subsection{The $\sigma$, Mg2, Fe diagnostic }

In Fig. \ref{fig_fe_sig}  the velocity dispersion and two iron indices
(Fe5270 and Fe5335)
are shown as a function of the Mg2 index,  separately for the sample of 
shell-galaxies and pair-members.

Shell-galaxies show a  narrower range of values with respect to
pair-members. Apparently the shell-sample does not contain low velocity dispersion
objects. Furthermore  all shell-galaxies lie above the {\sl universal}
$\sigma$ vs. Mg2
relation by Bender et al. (1993). On the contrary, the pair-members
nicely follow this relation.

The behavior  in the other index-index diagrams is quite
similar for the two samples even if they differ in some details. First
shell-galaxies tend to have both strong Mg2 and Fe indices at the same
time, whereas pair-galaxies have a broader distribution both in Mg2
and  Fe indices. Furthermore, in our sample, while the {\sl weak lined} galaxies 
(i.e. low values of metal indices), 
pair-members only, are almost consistent with the theoretical
prediction, {\sl strong lined} galaxies fall below it. 
In fact, it has long been known that (e.g. Worthey et al. 1994) the
relation between Fe indices and Mg2  tends to be flatter than that
traced by the evolutionary path of SSPs.
This fact has been interpreted as evidence of enhancement in
$\alpha$ elements with respect to the solar partition (perhaps
caused  by Type II super-novae contamination). We will come back later
to this topic.

The peculiar $\sigma$ vs. Mg2 diagram of shell-galaxies conforms to 
what is found for field galaxies with the fine structure index
$\Sigma$ of Schweizer
(1992) larger than 2 (Bender et al. 1993). However in our sample the
effect is more significant.

Among the galaxies showing fine structures,
shells are believed to be a signature of past strong dynamical interaction
(Barnes 1996).
If this  mechanism is responsible of the above displacement, several
explanations  are possible: (i)
the major effect of dynamical interaction  is on the
velocity dispersion and other structural quantities rather than on the
photometric properties of shell-galaxies; in such a case the object moves
vertically away from the universal $\sigma$ vs. Mg2 relation and maintains its
position in the index-index planes. 
(ii) As suggested by dynamical models of strong unbound encounters,
the central velocity
dispersion of the interacting objects remains unchanged and the main effect is a
burst of star formation altering only the spectro-photometric
properties, i.e. displacing them toward {\sl bluer, younger} values of
Mg2.
Most likely a combination of the two effects is at work.

Though far from being statistically complete, we argue that our sample could
sketch two consecutive phases of the accretion process: interaction and merging.
We note that our selection criteria are biased toward pair-objects obeying the
universal $\sigma$ vs.  Mg2 relation.

\begin{figure}
\psfig{file=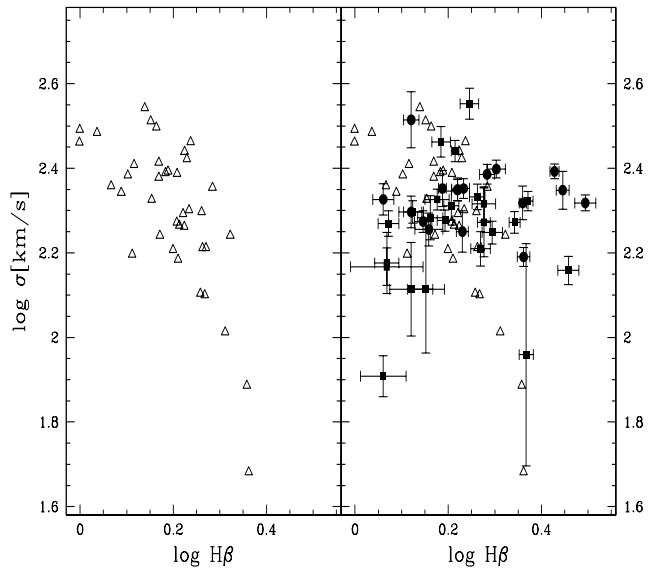,height=7.8truecm,width=8.7cm}
\caption{Left panel: correlation found by G93 between  velocity dispersion
and  H$\beta$  for   a sample of 40 {\sl normal} elliptical galaxies. Right
panel: the same but for shell- (full circles) and pair-galaxies (full
squares)   superposed to 
the {\sl normal} galaxies by G93. The error bars are also indicated  }
\label{fig_sig_hb}
\end{figure}

\subsection{ On the enhancement of $\alpha$ elements }

Assuming that Fe5270 and Mg2 are good indicators of the iron and
magnesium abundances $Z_{Fe}$ and $Z_{Mg}$, respectively, G93
and Worthey et al. (1992)  inferred that the [Mg/Fe] ratio changes with
the mass of galaxies. Giant galaxies are characterized by a high
[Mg/Fe] ratio, whereas ordinary galaxies by a lower, nearly solar ratio.  

Several scenarios have been proposed by G93, Worthey (1992), Worthey
et al. (1992) to explain the observed Fe vs. Mg2 relation and the
inferred over-abundance of Mg passing from ordinary to giant (bright)
early-type galaxies (see also Matteucci 1997 for a recent review of
the subject).

All of
these interpretations stand on the notion  that heavy elements (like Fe) are
predominantly generated by Type I super-novae,  whereas 
lighter elements (like O, Mg, Si) are produced by Type II super-novae. 
As Type I and II  SNe have progenitors with very different stellar
masses (intermediate, low-mass stars the former,  and massive stars the 
latter), the production of O, Mg, Si occurs much earlier than the bulk
production of Fe.

The goal is reached by supposing that star formation stops much
earlier in giant galaxies than in  compact ones. 
G93's suggestion is somehow supported by 
the simple interpretation of the distribution of early-type galaxies in
his sample on the H$\beta$ vs. [MgFe] plane (see below).
Furthermore, considering the velocity dispersion  $\sigma$ as a measure
of the galaxy mass, he argued that  massive galaxies appear to be
older than the low mass  ones.

Incidentally,
this opposes the standard galactic wind scenario proposed long ago
by Larson (1974) to explain the color-magnitude relation of
early-type galaxies.

It is worth noting that short-lived  star formation in  giant
(massive)  galaxies
with respect to that in the  compact (less massive) ones is not 
fully  supported  by major current theories of galaxy formation. 
In fact, in the gravitational collapse  scenario,
star formation  in giant elliptical galaxies is expected to last longer
than in the compact
galaxies, because in the former  the dynamical time scale is larger 
(about 3 times) than  in the latter. Conversely, in  the hierarchical
scenario, where big galaxies  are the result of several mergers, 
global star formation is expected to last for long periods of time.
In any case, both scenarios predict giant galaxies to be {\sl younger} than 
compact galaxies. 

To cast light on this intriguing affair, 
Borges et al. (1995) have recently provided a new empirical calibration
of the Mg2 index, that takes explicitly into account the
relative abundance of the Mg element.
They express the Mg2 index as a function
of effective temperature, gravity, [Fe/H] and [Mg/Fe].
Tantalo et al. (1998) included the new calibration of Borges et al. (1995) 
in their models of synthesis of stellar populations, and
they calculated the expected integrated Mg2 index for SSPs
of various ages and metallicities, assuming the same isochrone
for different abundance ratios at a given total metallicity. 
From their analysis of the G93 sample, younger 
galaxies seem to be moderately more metal rich and less
[$\alpha$/Fe] enhanced than the older ones (see figures 5 and 6
in Tantalo et al. 1998). The present study adopts  the ``standard'' 
fitting functions, based on solar abundance ratios, and thus it
cannot follow their detailed abundance analysis.
Nevertheless, 
their results will appear to be strengthened 
when a larger sample of galaxies is analyzed in the [MgFe] vs. H$\beta$
diagram (see next section).

\subsection{The $\sigma$ vs. H$\beta$ diagnostic}

According to Worthey (1992) and Bressan et al. (1996)
MgI, MgH and FeI indices are good metallicity indicators whereas
 H$\beta$  is more suited  to age determinations.

According to
G93, in the case of SSPs,  H$\beta$ is a good age indicator as it reflects 
the temperature of the  turn-off  stars. In galaxies, recent
experiments with composite stellar populations suggest that
H$\beta$ yields  a sort of {\sl mean age} of the stellar
populations, as it  tends to over-weight the age of
the young stellar component with respect
to the age of the bulk  stars (see e.g. Bressan et al.
1996).

The two panels  of Fig. \ref{fig_sig_hb} make evident the different
behavior  of our sample (right)
and  G93 sample (left panel) in the  $\sigma$ vs. H$\beta$  plane. G93 found
a correlation between $\sigma$ and H$\beta$  indicating that {\sl normal}
ellipticals, characterized by a shallower gravitational potential, are
predominantly young objects as indicated by their high value of H$\beta$
(see also Fig.~11 in Bressan et al. 1996).

In our sample  there are a number of galaxies  with
high $\sigma$ and H$\beta$ values at the same time. This is
particularly  true for the sample of shell-galaxies. This finding 
implies the existence
of a family of galaxies with deep gravitational 
potentials and relatively young ages.

This can be a consequence of the interaction experienced by these  galaxies that 
increases both  $\sigma$  and H$\beta$ (thus making them appear
younger).

Rampazzo et al. (1999a)  have shown that shell-galaxies seem actually  to belong to the 
family of giant galaxies, as
far as
their effective radius and surface brightness are concerned (see Fig.2
in Rampazzo et al. 1999a). In contrast,
pair-members have  a much wider distribution, going from normal to
giant galaxies. 

Following G93, we would 
expect that shell-galaxies (high $\sigma$ and high mass in turn)
are old stellar systems, in which the 
young component formed during a recent interaction and/or merger
episode is either too old  or too weak  to be able to affect  
the global Mg indices. The fact that the {\sl strong lined}
pair-members show the same behavior of shell-galaxies (i.e. deviation
from the Fe vs. Mg2 relation and likely  high [Mg/Fe] ratios) whereas the
{\it weak-lined} ones agree with the above relation and likely have 
{\sl normal} [Mg/Fe] ratio, leads us to argue that a much larger spread
in age ought to exist among pair-members as compared to
shell-galaxies. Checking this point requires the use of H$\beta$ as 
age indicator (see below).

As our sample is composed of objects in  interaction  and/or 
post-interaction stages,  the mean age of their stellar populations 
is less meaningful
with respect to that of the {\sl normal} ellipticals of G93. In fact, an
old  age for the {\sl giant} galaxies does not exclude the possible
presence of a young component formed in a recent (or ongoing) event
involving only a small percentage of the total mass of the galaxy.

\subsection{H$\beta$ vs. [MgFe] diagnostic  }

\begin{figure}
\psfig{file=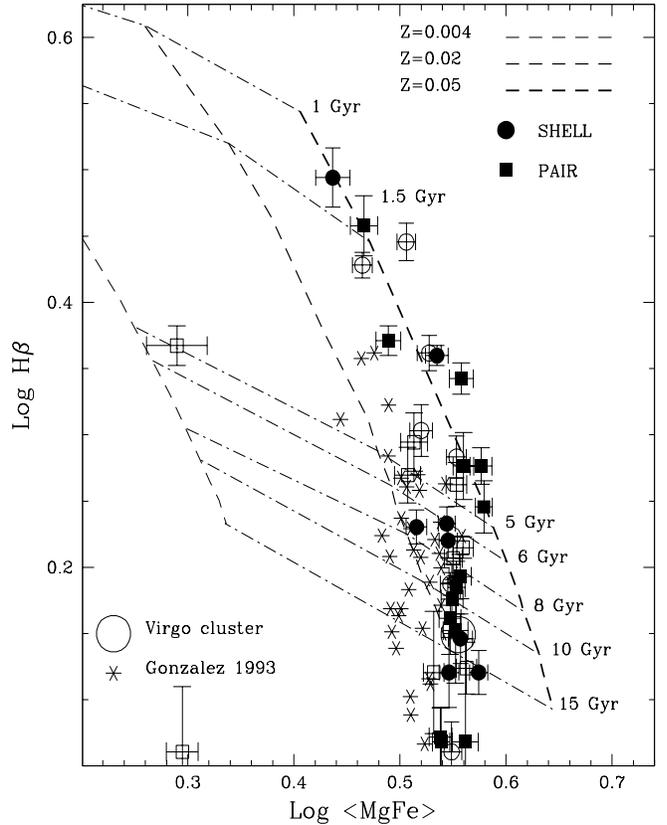,height=11.0truecm,width=8.8truecm}
\caption{The position of shell- (filled
dots) and pair-galaxies (filled squares) together with their
 1$\sigma$ error bars onto the [MgFe] - H$\beta$ plane.
Open circles and squares indicate shell galaxies and pair galaxies
(respectively) showing emission lines. 
The asterisks indicate the {\sl normal} elliptical galaxies by G93, 
for purposes of comparison. The large open circle
is  the average value for the 7 galaxies in the G93 sample members of
the Virgo.  Superposed to the data are the evolutionary paths (dashed
lines) of  SSPs 
with  different metallicity as indicated. The models go from the age
of 1.0 (top) to 15 Gyr (bottom).
Lines of constant age are also shown (dotted-dashed lines)  }
\label{fig_hb_mgfe}

\end{figure}

Fig. \ref{fig_hb_mgfe} presents the shell- and pair-galaxies of our
sample (dots and squares, respectively) 
in the 
[MgFe] - H$\beta$ plane and compares them with theoretical models. 

Three SSPs (dashed lines) with different chemical composition, i.e. [Y=0.250 Z=0.004],
[Y=0.280 Z=0.020]  and [Y=0.0.335 Z=0.05], are plotted as a function of the
age, and lines of constant age are also indicated (dotted-dashed).

In the same diagram we plot the G93 galaxies (asterisks) and the mean 
value for the early-type galaxies of the Virgo cluster (large open circle).
The two sets of data share the same properties but differ in some
important 
aspects. Pair- and shell-galaxies show a global metallicity  similar
to that of the {\sl normal} G93 galaxies, and no difference between the
shell-galaxies and pair-members appears in this diagram. At the same
time we notice that the H$\beta$ values for our sample extend up to
the top of the [MgFe] - H$\beta$ plane, where {\sl normal} galaxies are
not found.

There is a group of galaxies whose H$\beta$ is much lower  than predicted by
models
of old age (say 15 Gyr). These galaxies are likely to have H$\beta$
affected by contamination of the H$\beta$ emission line whose net effect
is to decrease the value of H$\beta$ by partially filling up the
corresponding absorption feature.

The last thing to note is that the distribution of galaxies in this
plane, of shell- and pair-objects in particular, is significantly steeper
than the path followed by SSPs. The locus of real galaxies stretches
almost vertically along the H$\beta$ axis and covers a narrow range
of [MgFe].

\section{Individual galaxies on the H$\beta$ vs. [MgFe] plane}

This section is dedicated to discussing in some detail 
the location of groups of galaxies on the H$\beta$ - [MgFe] plane
trying firstly  to understand the reason of their position 
 and secondly to decipher the underlying age.
Particular attention is paid  to some of the galaxies in the
sample, the shell-galaxies in particular. In fact for a few of them
independent estimates of the age of the last burst
of star formation based on morphological observations and dynamical
models can be found in the literature. Consistency between photometric and
dynamical age estimates would provide useful insight on the
connection between dynamical processes and star formation.

\subsection{Galaxies showing emission lines}
Some of the galaxies of our sample show clear emission lines.
In particular, among the pair galaxies, we find a detectable [OII](3727\AA) line
in 10 objects (in RR278a this line can be detected but
its measure is very uncertain for the high noise of the corresponding
spectrum around 3700\AA). A subsample of 3 out of 10 galaxies
(RR24a, RR24b and RR278a) shows also the [OIII](5007\AA) line
and the H$\beta$ emission.
Among  shell galaxies, the  [OII](3727\AA) line can be detected in 7 objects
4 of which (NGC~7135, NGC~1210, NGC~6958 and NGC~2945) show
also the [OIII](5007\AA) line, whereas the  H$\beta$ emission can be detected 
(even if measured with great uncertainty) only in NGC~7135. 

In Fig. \ref{fig_hb_mgfe} the four galaxies (3 pair objects and 1 shell galaxy)
showing H$\beta$ in emission are not shown (their H$\beta$ index
being negative).  

As far as the 
remaining galaxies
with measured [OIII](5007\AA) emission (NGC~1210, NGC~6958 and NGC~2945)
are concerned,
adopting the G93 correction they would shift by $\delta$LogH$\beta$=+0.27,
+0.67 and +3.4, respectively.

Given the high expected correction we have decided not to
display  these galaxies in Fig. \ref{fig_hb_mgfe}.   

For the galaxies  showing only [OII](3727\AA) line in emission (3 shell
galaxies and 7 pair members), we used open symbols, in order to stress that
the value of the H$\beta$ index can be affected by the emission contamination.
However, by considering that the threshold of [OIII](5007\AA) detection in our
spectra is  about 0.1\AA, we estimate that in these latter galaxies 
the correction to the observed
Log(H$\beta$)  is less than +0.15 dex.

As previously pointed out, no correction for emission  has been applied to 
the values of H$\beta$ shown in Fig. \ref{fig_hb_mgfe}.

\subsection{Galaxies with low [MgFe] values}

RR~317b and RR~62a have unusually low values of [MgFe]  and very 
different values of H$\beta$, very high in RR~317b and very low
in RR~62a. Furthermore, while RR~317b shows evidence of
emission [OII](3727\AA), no detection is found
in  RR~62a (Longhetti et al. 1998b).
Rampazzo et al. (1999a)  have shown that in the $\mu_e$ - R$_e$ plane both
galaxies  occupy  the same area as
{\sl ordinary} galaxies. In addition to this,  
RR~62a lies  outside the Hamabe-Kormendy (1987)
relation which holds for  bright galaxies. 
All this  implies that both galaxies are not giant objects.
According to Gorgas et al. (1993) 
 dwarf galaxies populate the left part of the H$\beta$ - [MgFe]
plane showing a metallicity lower than in giant ellipticals. 
In this context, the location of RR~62a and RR~317b can be
explained.

\subsection{Galaxies with high H$\beta$ values}

Bressan et al. (1996) investigated the evolutionary path in the
H$\beta$ vs. [MgFe] plane of a galaxy in which a recent burst of 
star formation
is superposed to a much older population.

As soon as the burst begins, the galaxy  runs away from the
locus of quiescence toward  the top of the  H$\beta$ - [MgFe]
plane following a nearly vertical path.
As soon as the burst is over, on a short time scale, the galaxy  goes
back 
toward its original position sliding along a path that is also nearly
vertical. A loop is performed.
See the numerical experiment described by Bressan et
al. (1996), in which a burst with  1\% of the galaxy mass engaged in the
star forming activity and 
duration of 10$^{8}$ years is calculated and displayed in the 
H$\beta$ vs. [MgFe] plane (Figs. 7 \& 8  in Bressan et al. 1996).
Since 
the burst intensity and  duration may vary from one case to another,
a large variety of loops are possible in the H$\beta$ vs. [MgFe]
plane. Stronger and longer
bursts induce wider  loops and longer recovery time
scales.

Looking at post-star-burst objects such as the shell-galaxies, they
are  expected to
closely follow the distribution of star-burster simulations and to
somehow depart from the region occupied by the {\sl normal} galaxies
of G93.
This could be the case for galaxies with very
high values of H$\beta$ (LogH$\beta >$ 0.35) 
present in our sample, namely NGC~813,
NGC~2865, NGC~5018, ESO~2400100b, ESO~1070040, RR~187b, RR101b. None of these
galaxies show emission lines in their spectra, apart from ESO~2400100b
that shows only the [OII](3727\AA) line. 
With  lower H$\beta$ (0.25 $<$ LogH$\beta <$ 0.35), 
but different position with respect to the G93 sample, there are also
NGC~1316, NGC~6776
(with [OII](3727\AA) emission detected in their spectra), RR~225a, 
RR~387b RR~409a, RR297a, RR101a and RR~405b. 

\underbar{\bf Pair-members}.  None of the pair-members in this area
show emission lines in the spectrum, apart from RR101a that
shows the [OII](3727\AA) line. All systems  they belong to
show signs of interaction at least in one of the members. 
From the morphological
signatures and the absence of emission lines we
suggest  that
these pairs have already passed through the strongest interaction phase and
 are now  recovering from a burst
that likely occurred   less than 1 Gyr ago.

\underbar{\bf ESO~2400100}. Longhetti et al. (1998a) showed that 
ESO~2400100 is
actually made up of two components separated by 5\arcsec\ and 200
km~s$^{-1}$. This feature is present in three independent spectra.
Probably what we are seeing is an ongoing merger accompanied by
star formation as suggested by the very high 
value of H$\beta$. The component ESO~2400100a falls, however, 
in the   area  of  {\sl normal} galaxies.

\underbar{\bf NGC~2865}.
This shell-galaxy is at the top of the [MgFe] vs. H$\beta$ plane. Our
data, interpreted at the light of Bressan et al. (1996) simulations,
suggest that a burst of star formation has occurred very
recently, probably  less than 1 Gyr ago.
This galaxy has been observed by 
Schiminovich et al. (1995), who derive from VLA images,
a correlation  between the HI distribution and the fine structures 
(shells, tails, and loops) hosted by the galaxy. The authors argue that   
the lack of HI gas in the center  could be explained by a burst of star
formation. 
The origin of the shells and the evolution of 
NGC~2865 remain in any case unclear. The question to be asked is why the [MgFe] -
H$\beta$ plane hints at a recent burst, while the correlation
between shells and stellar motions favors  $\approx$7 Gyr old merger.

\underbar{\bf NGC~1316}. 
The position of this galaxy in the H$\beta$ vs. [MgFe] diagram suggests
that it has undergone a recent burst of star formation. 
Very recently,
Mackie \& Fabbiano (1998) studied the evolution of gas and stars
in the optical and X-ray bands. They
emphasized the presence in NGC 1316 of a complex tidal-tail system,
that hampers the accurate
reconstruction of its past history (mergers).
Speculating about some observational indications which hint at 
an efficient conversion of HI into other gas phases,
Mackie \& Fabbiano (1998) suggest that a low-mass, gas-rich galaxy
could  have started  merging $\approx$ 0.5 Gyr ago.
A longer time scale is proposed by 
Schweizer (1980) to explain NGC 1316 as the product
of several mergers of gas-rich galaxies over the past $\approx$ 2 Gyr.

\subsection{Galaxies with normal [MgFe] and  H$\beta$ values}

In the area occupied by the majority  of the G93 galaxies, in particular those
which are members of the Virgo cluster, we find both shell-galaxies and 
pair-members. The age of the bulk stellar population in these galaxies can
be estimated 
as old as 10-15 Gyr. If a star-burst phenomenon has ever occurred, its 
signatures on the stellar populations of the galaxies located
in this area of the diagram have already faded away.
Shell-galaxies falling in this region 
(NGC~1549, NGC~1553, IC~5105, NGC~6849) 
suggest that shells are long-lasting morphological features,
which can be detected also after star formation signatures have
disappeared.

\begin{figure*}
\psfig{file=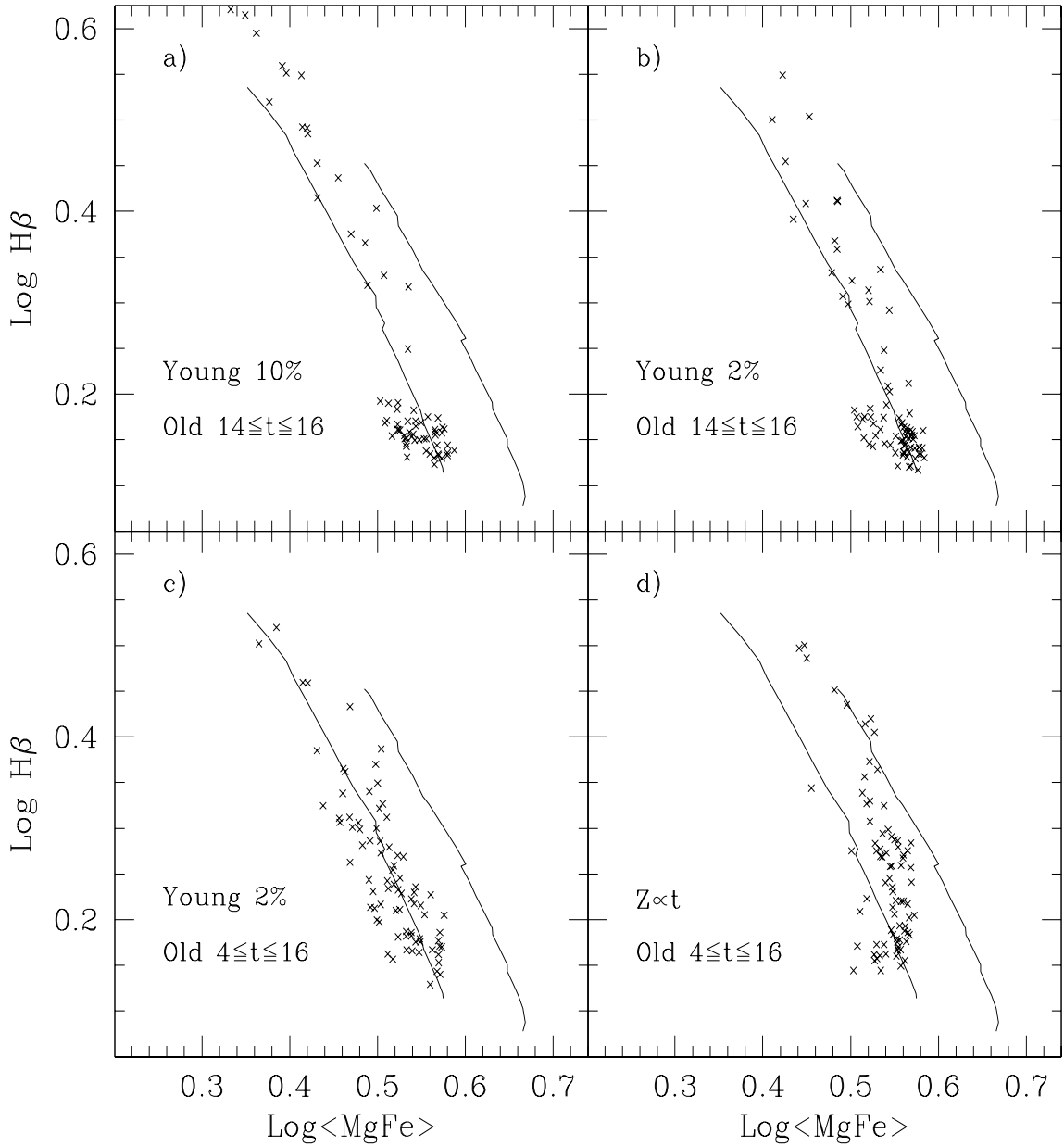,height=20.0truecm,width=18truecm}
\caption{Simulations of 80 galaxies in the H$\beta$ vs. [MgFe]
plane. The age of the burst is randomly chosen between 0.1 Gyr and 1 Gyr  
 for 20 galaxies (simulated pair-members) and the bulk age for all remaining
objects (60 in total). 
The fraction of the mass stored in the young component is randomly chosen 
from zero to either 10\% -panel (a)- or 2\% -panels (b), (c) and (d)-.
The age of the bulk (old) component is randomly fixed between the values
indicated in the corresponding panels. In panels (a), (b) and (c) 
the metallicity
is allowed to randomly vary from Z=0.015 to Z=0.025. In panel (d) the metallicity is
allowed to linearly increase with age from Z=0.015 to Z=0.04.
Random errors  $\Delta$Log([MgFe])$\leq$0.011 and
$\Delta$Log(H$\beta$)$\leq$0.02 are applied to model indices to better 
simulate the data. Finally, the solid lines are  two SSPs with Z=0.02 (left)
and Z=0.05 (right). } 
\label{fig_hb_mg_sim}
\end{figure*}

\section{More on the H$\beta$ vs. [MgFe] plane}

Starting from the idea of Bressan et al. (1996) that the distribution
of galaxies in the H$\beta$ vs. [MgFe] plane may reflect secondary
episodes of star formation superposed to an old stellar component,
we present here the results of simulations designed to clarify this point.
In the analysis below, we will neglect any  effects due to 
[$\alpha$/Fe]. As a matter of fact, 
first no calibration accounting for
different [Mg/Fe] ratios is available for the Mgb index (in analogy to
that of Borges et al. (1995) for the Mg2 index). Furthermore,
any increase in the [Mg/Fe] is expected to be associated 
with a decrease in [Fe/H], and the two effects likely parallel each other
in final determination of  [MgFe].
Therefore, [MgFe] can be safely considered
as a good, global indicator of metallicity (see also G93).

We have constructed models aimed at predicting the statistical distribution
of 80 post-burst galaxies, and to reproduce
the observational one displayed in Fig.~\ref{fig_hb_mgfe}. 
Following the selection criteria explained in
the previous section, the observational sample to be matched is 
composed of about 20 shell galaxies, 20 pair members
and 40 ``normal'' galaxies from G93. 

We simplify the complex star formation history of an early--type galaxy 
to a bulk population made of old stars on which a more recent burst of
activity with different age and intensity 
is   superposed. The old stars are in turn
represented by
a SSP whose age is randomly chosen between $T_1$ and $T_2$,
parameters in the models.
We are aware that this oversimplified scheme neglects the
complex star formation histories of real galaxies and
the ensuing mixture of stars with different chemical composition.

Integrated indices of SSPs, resulting from multi-population models,
have been investigated by Idiart et al. (1996) and Bressan et al. (1996). 
In particular, the complex chemical modeling used by
Bressan et al. (1996) suggests that field early--type galaxies are
confined within  a narrow range of metallicities, around the mean value.
For the sake of simplicity, we will use the mean metallicity 
(despite the physical process by which galaxies get enriched 
in metals, e.g. closed-box, infall,  etc.) to rank galaxies as a function 
of this parameter.

The young stellar component, formed during the recent burst
of star formation, is represented by a SSP whose age is randomly varied 
between 0.1 Gyr and  $T_3$. The value of $T_3$
has been fixed to 1 Gyr for 20 models 
in order to represent the 20 pair members. These latter in fact 
seem to be characterized by a very young stellar component 
generated by the ongoing dynamical interaction
(Longhetti et al. 1999).

For the remaining 60 objects, $T_3$ is allowed to arbitrarily vary from young 
to old ages (up to $T_2$). 
The strength of the secondary
burst, represented by the percentage of the total mass turned into 
stars, is randomly chosen 
between 0 (no secondary stellar activity) and $f$, a parameter
of the models. Since we are interested in guessing the minimum
threshold above which the secondary episode has a sizable effect on 
the line strength  indices, we will consider only the case
in which the secondary episode involves a minor fraction of the
galaxy mass.  

The metallicity $Z$ of both stellar components
is randomly varied from Z=0.015 (75\% of the solar value) to Z=0.025
(1.25 the solar value). We have also calculated a set of simulations
in which the metallicity is supposed to linearly increase
from Z=0.015 to Z=0.04 (twice the solar value) over the time interval 
$T_1$--$T_2$ .  

Finally,  random errors  $\Delta$Log([MgFe]) $\leq$ 0.011 and
$\Delta$Log(H$\beta$) $\leq$ 0.02 are applied to the model
indices 
to better simulate observational data.

In Fig.~\ref{fig_hb_mg_sim} we show an example of the 80 simulated 
galaxies for 4 different models. The aim is
to highlight the effects of the underlying basic parameters, namely
the epoch at which the bulk stars have been formed, 
the strength of the more  recent burst episode, if any, and its age.

In panels (a) and (b) galaxies are conceived as old, nearly coeval systems,
their population being approximated by a single SSP with age between 14 and
16 Gyr; this corresponds to the current view of elliptical galaxies in rich
clusters (Bower et al. 1998).
The maximum intensity of the superposed burst amounts to  
10\% and 2\% of the total
mass, panels (a) and (b) respectively. 

Inspection of the results shown in  panel (a)  
indicates that  model galaxies with a burst engaging a percentage 
of the total mass up to 10\% 
predicts too high values of H$\beta$ and 
too many {\it young} objects. In contrast, the models of panel (b) 
span the range of H$\beta$ values indicated by the observations.
In both cases, however, 
the expected distribution of objects with respect to the 
H$\beta$ index is at variance with the observational one.
Indeed models of this type predict a bimodal distribution, whereby
the old galaxies (those for which the burst is almost as old as
the bulk of their stellar populations) clump together in the 
lower portion of the diagram,  whereas the ``young'' objects 
(those with very young
bursts) form a tail extending to high values of H$\beta$. 
The real clump is even narrower than displayed if one considers the
effect of the simulated errors.

The tail of ``young'' models 
agrees quite well with the observations, thus suggesting that 
the upper part of the diagram is populated by objects
which experienced a recent burst of star formation of minor entity 
(less than 2\% of the total mass).

The burst alone
cannot, however,  explain the smooth distribution observed at 
low values of H$\beta$.
The reason for it is that the index H$\beta$ of a composite 
stellar population (2\% of the mass is ``young'' stars and the remaining 98\%
in old stars), soon after the maximum excursion toward high values of 
H$\beta$, fades out  very rapidly as the young population ages.  Therefore
catching a galaxy at intermediate values is highly improbable. 

Slowing down the evolutionary rate of the ``young'' component
by increasing the
percentage of mass involved in the young burst produces an uncomfortably  
large fraction of objects in the upper part of the diagram, e.g.  panel (a).

Since a sort of fine tuning between the old and young component is  not 
easy to understand, the viable alternative is that the old component spreads 
over significantly longer periods than assumed in the above simulations.
To this aim, we present the models shown
in panels (c) and (d). The {\it old} population in these models has an 
average age that spreads over a significant fraction of the Hubble time.  
This is meant to
indicate that either the object has been growing for such a long time
with a low star formation rate, or that its major star formation activity 
was not
confined to an early epoch. The young component is allowed to occur as
in the simulations of panel (b). It is immediately  evident 
that the new models much better reproduce the distribution
of galaxies all over the range of H$\beta$.

However, problems remain if the metallicity is 
randomly chosen. 
Indeed, the model galaxies always tend to follow 
the path of an SSP,  in these simulations the SSP with 
Z=0.02 as shown in panel (c).  As already anticipated, real galaxies 
seem to follow a 
path steeper than that of  SSP. 
We consider this point as  evidence of an underlying relation 
between the age and metallicity of the bulk population of stars. 
 Panel (d) shows our final experiments, where
the average metallicity of the bulk stars 
has been supposed to linearly increase with age from  Z=0.015 to Z=0.04.
The broad  range of metallicity is required by the large scatter 
in Log([MgFe]).  If this suggestion is sounded, it would imply 
that young early type galaxies in the field are on average
more metal-rich than old systems, with an average rate of metal-enrichment
$\Delta$Log(Z)/$\Delta$Log(t)$\simeq$-0.7.
If confirmed, the vertical distribution of galaxies
in this diagram is therefore {\sl the trace of a global metal enrichment  
taking place in galaxies during all the star forming episodes}.

\section{Summary and conclusions}

In this  paper firstly we have compared the line strength indices 
for SSPs 
that one would obtain  using different calibrations in
literature, namely Worthey (1992), Worthey et al. (1994), 
Buzzoni et al. (1992, 1994) and Idiart et al. (1995).
Secondly, with the aid of the Worthey (1992) and Worthey et al. (1994)
calibrations, the  sample of shell- and pair-galaxies
by Longhetti et al. (1998a,b), and the sample of G93
for {\sl normal} elliptical galaxies have been systematically
analyzed, looking at the position of all these galaxies in various
diagnostic planes. The aim is to
cast light on the star formation history that took place in these
systems with particular attention to those (shell- and pair-objects)
for which the occurrence of dynamical interaction is evident. Finally,
from comparing normal to dynamically interacting galaxies we attempt
to understand the reasons for their similarities and differences.

The results of this study can be summarized as follows:

(1) The various calibrations for the line strength indices as a function
of basic stellar parameters (effective temperature, gravity and
metallicity) lead to quite different results.
Specifically, the Buzzoni et al. (1992, 1994) calibrations for Mg2
and Fe5270  agree with those by Worthey (1992) and
Worthey et al. (1994) only for SSPs older than 3 Gyr.
For H$\beta$  the agreement is also  good if one  excludes 
all ages younger than about 1 Gyr. 
Idiart's et al. (1995) calibrations can be compared with
the others only for a limited range of SSP ages. 
For the Mgb and Mg2 indices we find a roughly constant offset,
that could be attributed to different properties 
(g.e., metallicity, gravity) of the calibrating sample of stars. 
For the purposes of this study we have
adopted Worthey (1992) and Worthey et al. (1994) as the reference
calibrations.

(2) The  comparison of the Mg and Fe indices 
(specifically Mg2, Fe5270 and
Fe5335) and the velocity dispersion $\sigma$
of normal, shell- and pair-galaxies suggests  firstly a different
behavior of shell-galaxies with respect to pair-objects and secondly 
 that strong-lined
galaxies  are likely to have  super-solar   [Mg/Fe]
abundance   ratios. Various kinds of star formation histories leading
to
super-solar [Mg/Fe] are examined at the light of current understanding
of the mechanisms of galaxy formation. None of these is however able to
give a self-consistent explanation of the [$\alpha$]-enhancement
problem.

(3)  The same galaxies are analyzed in the
H$\beta$ vs. [MgFe] plane and compared to the G93 set of
data.
The  shell- and pair-objects have the same distribution in this 
diagnostic plane as the {\sl normal} galaxies even if they
show a more pronounced tail toward high H$\beta$ values.

(4) The H$\beta$ vs. [MgFe] plane is divided in several sub-regions
carefully inspected in order to look for  
all plausible causes that would justify the positions
of the galaxies.

(5) As shell- and pair-galaxies share the same region of the  H$\beta$
vs. [MgFe] plane occupied by {\sl normal} galaxies, we suggest that a
common physical cause is at the origin of their distribution.
The star formation history  in these objects is investigated with the 
aid of very simple galaxy models. 

We find that the tail at high H$\beta$ values can be 
ascribed to secondary bursts of star formation which, in the case of
shell- and pair-galaxies, can be easily attributed to
interaction/acquisition events whose signatures are well evident in
their morphology. 

(6) A typical model where the burst is superimposed to an otherwise
old and coeval population is however not able to reproduce the smooth
distribution of galaxies in the H$\beta$ vs. [MgFe]
plane. This kind of model would predict an outstanding clump at low 
H$\beta$ values, contrary to what is observed.
Models in which the bulk of the star formation happened over
a significant fraction of the Hubble time ( 4 Gyr $\leq$ t$_{old}$ $\leq$ 16 Gyr)
better match the observed diagram. 

(7) In this context, the peculiar, smooth and almost vertical distribution of
galaxies (normal, shell- pair-objects) in the H$\beta$ vs. [MgFe]
plane is interpreted as the trace of the increase of the {\sl average}
metallicity accompanying all star forming events. This could be the signature
of a metal enrichment happening on a cosmic scale.

(8) Although deciphering the position of galaxies in the H$\beta$ vs
[MgFe] plane to infer the age of the constituent stellar populations
is a difficult task due to  possible blurring caused by the secondary
stellar activity, still we may draw some conclusions for the duration
of the shell phenomenon. Specifically, since shell-galaxies can be
found in the same region of old normal galaxies, we may say that shells
are a morphological feature able to persist for long periods of time,
much longer than the star forming activity that likely accompanied
their formation. Among current dynamical models in which
shell-structures can be formed, the {\sl weak interaction} mechanism
of Thomson \& Wright (1990) and Thomson (1991) naturally predicts
long-lived shells without particular hypotheses on the type of
encounters.

\begin{acknowledgements}
ML acknowledges the kind hospitality of the  Astronomical Observatory of
Brera (Milan) and Padua during her PhD thesis and the support 
by the European Community under TMR grant ERBFMBI-CT97-2804.
CC wishes to acknowledge the friendly hospitality and
stimulating environment provide by MPA in Garching where this paper
has been completed during leave of absence from the Astronomy
Department  of the Padua University.  This study has  been financially
supported by the European Community
under TMR grant ERBFMRX-CT96-0086.
\end{acknowledgements}

\end{document}